\documentclass[%
reprint,
 amsmath,amssymb,
 aps,
 prl,
]{revtex4-1}

\usepackage{graphicx}

\begin{document}

\preprint{APS/123-QED}

\title{A method to approximate cluster integrals and 
\\ reproduce subcritical isotherms for model and real gases}

\author{M. V. Ushcats}
\email{mykhailo.ushcats@nuos.edu.ua}
\affiliation{Admiral Makarov National University of Shipbuilding, 
 9, Prosp. Heroes of Ukraine, Mykolayiv 54025, Ukraine}%

\author{L. A. Bulavin}
\affiliation{Taras Shevchenko National University of Kyiv, 
 2, Prosp. Academician Glushkov, Kyiv 03680, Ukraine}%


\date{\today}

\begin{abstract}
A method is proposed to approximate the unlimited subcritical set of Mayer`s reducible cluster integrals (i.e., the power coefficients of the known virial expansions for pressure and density in powers of activity) on the basis of information about several irreducible integrals (virial coefficients) and the saturation activity. For the Lennard-Jones fluid, the first four irreducible integrals are actually sufficient to reproduce gas isotherms (including the flat phase-transition region) with high accuracy at temperatures between the triple and critical points. At low temperatures, even the usage of the first-order irreducible integral only (second virial coefficient) yields almost the same accuracy that, in turn, makes the method applicable to the real fluids with unknown interaction potential. In particular, the calculated water isotherms are in good agreement with experiments.
\begin{description}
\item[PACS numbers] {05.20.Jj; 05.70.Ce; 05.70.Fh; 51.30.+i; 64.10.+h; 64.60.-i; 64.70.F}

\end{description}
\end{abstract}

\maketitle


Today, it seems that we know a lot about the first-order phase-transitions in simple or complex substances \cite{Hill,Binder} and, sometimes, we even can analytically describe a number of specific complicated effects on the phase-transition parameters \cite{Huber2015,Bulavin2018,Lazarenko}. However in essence, our knowledge remains phenomenological, simulation-based, empirical or half-empirical rather than directly grounded on some pure theoretical relations between the macroscopic behavior of matter and its actual microscopic features. As to the gas~-- liquid transition, the Lee~-- Yang approach \cite{LeeYang} to define the partition function zeroes (on a certain complex plane) had been the sole example of such pure theoretical advances for a long time (here we do not account for surely successful but non-realistic mean-field approximations \cite{Kac2,Lebowitz2}), and even then it allowed only the definition of phase-transition parameters for the specific two-dimensional lattice-gas model without an accurate description of single-phase states. The inadequate mathematical behavior of the well-known virial series \cite{Mayer} in dense states had been a serious challenge in physics for decades and widely studied as "the problem of the virial series divergence" \cite{Groeneveld1962,Ruelle1963,Lebowitz1,Briano,Martynov}.

Recent studies have closed the most disputable issues in the modern statistical theory of condensation \cite{PRL,JML,PRE4,PRE5,PRE6}. A new approach \cite{PRL,JCP1,PRE1,Bannur} to express Mayer`s cluster expansion \cite{Mayer} in terms of irreducible cluster integrals $ \{\beta _k \left( T \right) \}$ indicated the beginning of condensation exactly at density $\rho _S$ where the isothermal bulk modulus of virial expansions vanishes:

\begin{equation}
\sum\limits_{k \ge 1} {k{\beta _k}\rho _S^k}  = 1. \label{eq:1}
\end{equation}

Beyond this saturation density, $\rho _S$ in Eq.~(\ref{eq:1}), the correct equations in terms of irreducible integrals $\{ \beta _k \}$ yield the constant pressure at $\rho > {\rho _S}$. Studies \cite{PRE4,Pramana} of Mayer`s cluster expansion in terms of reducible cluster integrals $\{ b_n  \left( T \right) \}$, i.e., the virial expansions for pressure and density in powers of activity (fugacity) $z$:
\begin{equation}
\left. \begin{array}{l}
\frac{P}{{{k_B}T}} = \sum\limits_{n = 1}^\infty  {{b_n}{z^n}} \\
\rho  = \sum\limits_{n = 1}^\infty  {n{b_n}{z^n}} 
\end{array} \right\}, \label{eq:2}
\end{equation}
completely confirmed the above-mentioned conclusions: when the activity in Eq.~(\ref{eq:2}) reaches quantity
\begin{equation}
{z_S} = {\rho _S}\exp \left( { - \sum\limits_{k \ge 1} {{\beta _k}\rho _S^k} } \right) \label{eq:3}
\end{equation}
the series for density in Eq.~(\ref{eq:2}) diverges that yields the jump of density (from the $\rho _S$) at strictly constant pressure and obviously constant chemical potential. Subsequent studies of Mayer`s expansion with regard for the volume-dependence of cluster integrals \cite{PRE5} or on the basis of the lattice-gas "hole~-- particle" symmetry \cite{PRE2,PRE3,UPJ4,UPJ5,PRE6} as well as independent studies of $N P T$-ensemble \cite{PRErus} have additionally proved the fact that condensation begins at the $\rho _S$ point defined in Eq.~(\ref{eq:1}).

Thus, the rigorous statistical theory of condensation may now be regarded as existing. Unfortunately, despite its certain qualitative successes such theory remains almost useless in practical applications. Indeed, any accurate quantitative application of Eqs.~(\ref{eq:1}), (\ref{eq:2}), (\ref{eq:3}) (as well as the above-mentioned approaches to define the partition function zeroes \cite{LeeYang, PRErus}) to a certain substance requires calculation of thousands~(!) cluster integrals. Even for the simplest realistic models of intermolecular interactions, the calculation of virial coefficients up to eighth-sixteenth orders has turned to be an extremely laborious computational problem \cite{KofkePRL,UPJ2,Kofke2015} and, for the orders of thousands, such calculations remain technically impossible today (not saying about real substances where the interaction potential may be unknown exactly).

There have been some attempts to predict the behavior of high-order irreducible integrals approximately at some limited region of subcritical temperatures \cite{JCP3, Kofke}, but they have not changed the situation cardinally. Those extrapolations made the theoretical isotherms qualitatively similar to the real ones (namely, the discontinuity of the isotherm tangent was observed at the saturation point) though the theoretical saturation parameters still were far from the empirical data (at low temperatures, they differ in three-ten (!) times at least) that has led to some disappointment and even doubts about the cluster-based approach: it was hard to realize how the virial power series may yield a "sudden" phase transition almost from ideal-gas states at low temperatures without any additional modification of such equations (like the Maxwell construction).

In this Letter, a new method is presented which allows approximation of the unlimited subcritical set of cluster integrals for various substances in a manner that fundamentally improves the accuracy of Mayer`s cluster expansion and thermodynamic functions which can be expressed in terms of cluster integrals. The key element of the method, which cardinally distinguishes it from others, is the approximation from infinite to low orders in addition to the conventional direction from low to high orders. For a studied substance, this method requires knowing only a few first irreducible integrals or virial coefficients (to guarantee the adequate gaseous phase behavior) and empirical quantity of the phase-transition activity (which guarantees the correct assymtotics of high-order integrals).

It is important to note that any certain (even limited) set of irreducible integrals $\{ \beta _k \}$ always constructs the infinite set of reducible integrals $\{ b_n \}$. One of the simplest algorithms of such construction has been presented recently in recursive form \cite{PRE4}:

\begin{equation}
{b_n} = \frac{A_{{n},{n - 1}}}{n^2}, \label{eq:4}
\end{equation}
where
\[
{A_{n,i}} = n\sum\limits_{k = 1}^i {\frac{k}{i}{\beta _k}{A_{{n},{i - k}}}}; \; \;\;\;{A_{n,0}} = 1.\]

In fact, the infinite set of reducible integrals may be constructed of only the first order irreducible integral, $\beta _1$, (i.e., the second virial coefficient, $ {B_2} = - {\beta _1} / 2 $). Some set of several irreducible integrals (let us hereafter denote the maximum order in this set as $k_m$) can well define the difference between real and ideal gases only at states beneath the saturation point, but such incomplete $\{ \beta _k \}$ set is insufficient to accurately define the saturation point itself, i.e., the $\rho _S$ density in Eq.~(\ref{eq:1}) and $z_S$ activity in Eq.~(\ref{eq:3}): in the resulting infinite set of reducible integrals defined by recursive relation~(\ref{eq:4}), only the integrals of orders up to ${n_{m}} = {k_{m}} + 1$ are as accurate as the initial irreducible integrals, but all the others (i.e., the high-order integrals which are really responsible for the divergence~-- phase transition) are rather inaccurate and, hence, produce the divergence far from the true phase transition that, in turn, makes any practical application of the theory problematic.  

Here we should draw attention to one very important mathematical fact, which has been established recently \cite{Pramana}: the asymptotic behavior of high-order reducible integrals constructed on the basis of a certain set of irreducible ones (no matter of its accuracy and completeness) always corresponds to convergence radius~(\ref{eq:3}) calculated by using the same set of irreducible integrals in strict accordance with the Cauchy-Hadamard theorem \cite{Hadamard} (this behavior is illustrated in Fig.~\ref{fig:1}):

\begin{figure}
\includegraphics{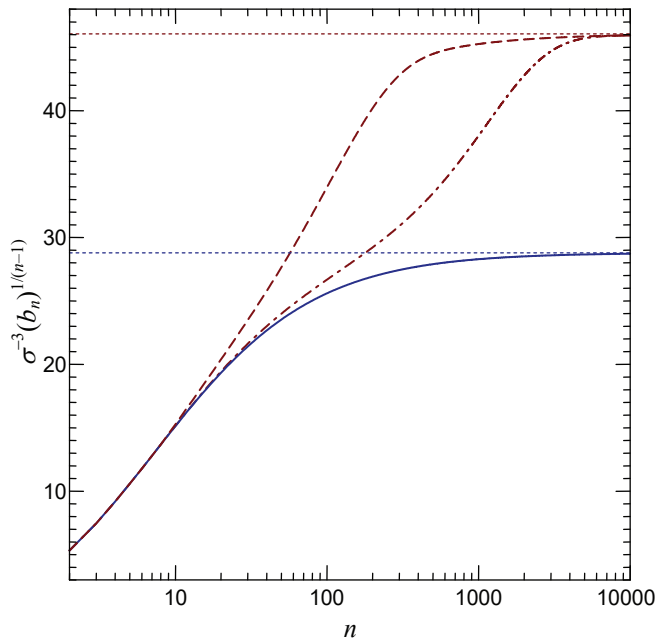}
\caption{\label{fig:1}Reducible integrals, $\{ b_n \}$, for the Lennard-Jones fluid ($ {k_B} T / \varepsilon = 1.0 $) calculated on the basis of the first seven irreducible integrals (solid curve) and integrals, $\{ b_{n}^{emp} \}$, approximated in Eq.~(\ref{eq:6}) with the $n_0$ parameter of $100$ (dotted curve) and $1000$ (dash-dotted curve). The horizontal lines show the reciprocals of the corresponding divergence activities: $z_S$ (bottom line) and $z_{S}^{emp}$ (top line).}
\end{figure}

\begin{equation}
\mathop {\lim }\limits_{n \to \infty } { b_n } = z_{S}^{ - {{\left( {n - 1} \right)}}} \label{eq:5} .
\end{equation}

Such mathematical feature has put the core idea of the proposed method to improve the accuracy of reducible integrals: if we have the true convergence radius (say its empirical value $z_{S}^{emp}$ known from some natural experiments or numerical simulations) we can confidently define the true values of high-order reducible integrals ($n \gg n_{m} $) which were evaluated inaccurately by using the limited $\{ \beta _k \}$ set. The simplest method to implement such correction is just to replace the inaccurate integrals by their true asymptotic values beginning at some order high enough, but this approach involves the uncertain choice of the order to begin the replacement and, moreover, it would yield the corresponding unrealistic jump instead of the originally smooth transition from low to high orders (the jump at the mentioned arbitrary chosen order). To avoid such jump within the interval of "middle-order" integrals (of orders transitional from the accurate low-order integrals to high-order integrals, which are also accurate now), we propose the following function to transform (or smoothly scale) the initial set of reducible integrals, $\{ b_n \}$, into a new empirically-based set, $\{ b_{n}^{emp} \}$:

\begin{equation}
b_{n}^{emp} \! = \! \left\{\begin{matrix}
b_{n}; & n\leq  n_{m}\\ 
b_{n} \! \left[ \! \left( \! \frac{z_{S}}{z_{S}^{emp}} \! \right)^{1-\exp\left( \! \frac{n_{m} - n}{n_{0}} \! \right)} \! \right]^{n-1}; & n> n_{m}
\end{matrix}\right. . \label{eq:6}
\end{equation}

Due to the additional exponent,
$ 1-\exp\left( -\frac{n-n_{m}}{n_{0}} \right) $,
and the $n_{0}$ constant there, the low-order integrals ($n_{m} < n \ll n_{0}$) close to the last correct one ($b_{n_{m}}$) also remains almost non-scaled while the others ($n \gg n_{0}$) are scaled so that their asymptotics in Eq.~(\ref{eq:5}) now corresponds to the correct convergence radius ($z_{S}^{emp}$ instead of $ z_S$). The choice of the $n_{0}$ constant (actually, it may vary from the values less than 1 to thousands and even more) regulates the scaling smoothness (from the complete scaling with a jump after the $n_{m}$ order to the almost initial non-scaled set of integrals) and the effect of this choice is considered in further paragraphs of the Letter. Actually, the usage of function~(\ref{eq:6}) is not a strict requirement: there can be a lot of possibilities for some reasonable transition from the low-order integrals (responsible for the correct gaseous phase behavior) to high-order integrals (responsible for the condensation at the correct value of activity).
 
To test the method, the Lennard-Jones fluid \cite{LJ1, LJ2} was chosen as a simple realistic model of intermolecular interactions which is widely used in theoretical studies and numerical simulations \cite{Kofke}. As a rule for Monte Carlo simulations, a gas~-- liquid coexistence point is defined by the equality of chemical potential for both phases and, hence, the phase-transition chemical potential is often presented in the results of such simulations. In Ref.~\cite{Lotfi}, this chemical potential is in the reduced form, $
\mu ^{*} = \frac{\mu - \mu _{id}}{k_{B} T} + \ln \left( \rho \sigma ^{3} \right) $, where $ \mu _{id} $ is the ideal-gas chemical potential at the same temperature and density and $\sigma $ is the distance parameter of the Lennard-Jones potential.

As to the activity used in Mayer`s cluster expansion \cite{Mayer}, $
z = {\lambda ^{ - 3}}\exp \left( {\frac{\mu }{{{k_B}T}}} \right) $ (in density units and in terms of the de Broglie wavelength, $\lambda  = h/\sqrt {2\pi m{k_B}T} $), its empirical phase-transition quantity may easily be transformed to the following form:
\[
z_{S}^{emp} = \sigma^{-3} \exp \left(  \mu ^{*} \right) , 
\]
taking into account that the ideal-gas activity is always equal to density ($ z_{id} = {\lambda ^{ - 3}}\exp \left( {\frac{\mu _{id} }{{{k_B}T}}} \right) = \rho $).

By using each empirical value of $\mu ^{*}$ presented in Ref.~\cite{Lotfi} for reduced temperatures (${{k_B}T} / \varepsilon $, where $ \varepsilon $ is the well-depth parameter of the Lennard-Jones potential) from $0.7$ to $1.3$ (almost from the triple point to the critical one), the set of reducible integrals, $\{ b_{n}^{emp} \left( T \right) \}$, was approximated in Eq.~(\ref{eq:6}) to the orders of $10000$ on the basis of the first seven irreducible integrals known for the Lennard-Jones model \cite{Kofke2015} and then the isotherms of Eq.~(\ref{eq:2}) where reproduced.

\begin{figure}
\includegraphics{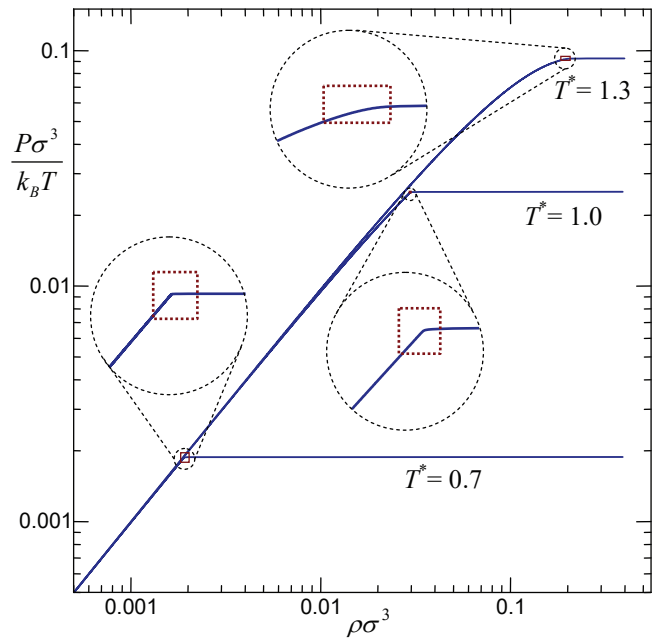}
\caption{\label{fig:2}Isotherms of Eq.~(\ref{eq:2}) with the set of $10000$ reducible integrals, $\{ b_{n}^{emp} \}$, approximated in Eq.~(\ref{eq:6}) ($n_{0} = 500$) on the basis of the first seven irreducible integrals and empirical data on the saturation chemical potential for the Lennard-Jones fluid at different reduced temperatures ($ T^{*} = {k_B} T / \varepsilon $). The dotted rectangles show the saturation parameters (within their uncertainties) from numerical simulations \cite{Lotfi}.}
\end{figure}

The first feature observed in the test calculations which needs special emphasizing is that, for each $\mu ^{*} \left( T \right)$ listed in Ref.~\cite{Lotfi}, the divergence of activity series (i.e., the density and pressure where the theoretical isotherm has the tangent jump corresponding to the saturation point) always takes place strongly within the uncertainties of the saturation density-pressure parameters presented in the same Ref.~\cite{Lotfi} (for three temperatures, this consistency is well illustrated in Fig.~\ref{fig:2}). With regard for the obvious approximate nature of the considered approach, such excellent accuracy has turned even somewhat surprising for the authors of the method. Although the studied temperature interval may seem not too wide the saturation pressure and density vary by two orders within this interval that forced to use the logarithmic scales in Fig.~\ref{fig:2}.

A change in the number of reducible integrals (i.e., the power coefficients in Eq.~(\ref{eq:2}) which must actually form the infinite sequence) from $1000$ to $10000$ slightly affects the isotherms at the vicinity of the saturation point only: the more integrals are taken into account in Eq.~(\ref{eq:2}) -- more sharp is the tangent jump there. Moreover, varying the $n_{0}$ parameter of Eq.~(\ref{eq:6}) in wide ranges from a number of dozens to thousands (see Fig.~\ref{fig:1}) has an extremely small effect on the resulting isotherms: for the isotherms presented in Fig.~\ref{fig:2}, the corresponding changes are absolutely invisible in the scales of this Figure.

An additional important feature of the presented approximation is the relatively fast convergence with the increase of the initial set of irreducible integrals, $\{ \beta _k \}$. Even at temperatures close to the critical one (for the Lennard-Jones fluid, $ T_{cr} \approx 1.32 \varepsilon / {{k_B}}$), the first four irreducible integrals, $\{ \beta _{1}, \beta _{2}, \beta _{3}, \beta _{4} \}$, are actually enough to construct the $\{ b_{n}^{emp} \}$ set which yields the isotherms almost indistinguishable from those obtained on the basis of seven irreducible integrals (see the top isotherm in Fig.~\ref{fig:2}) and the integrals of the fifth, sixth, seventh and even higher orders \cite{Kofke2015} have no significant effect on the results. As to the considerably lower temperatures, there the requirements for the number of initially known irreducible integrals become even much softer. Actually, the usage of only the first irreducible integral (i.e., only the second virial coefficient) instead of the seven integrals almost does not affect the resulting low-temperature isotherms (see the two bottom isotherms in Fig.~\ref{fig:2}). 

\begin{figure}
\includegraphics{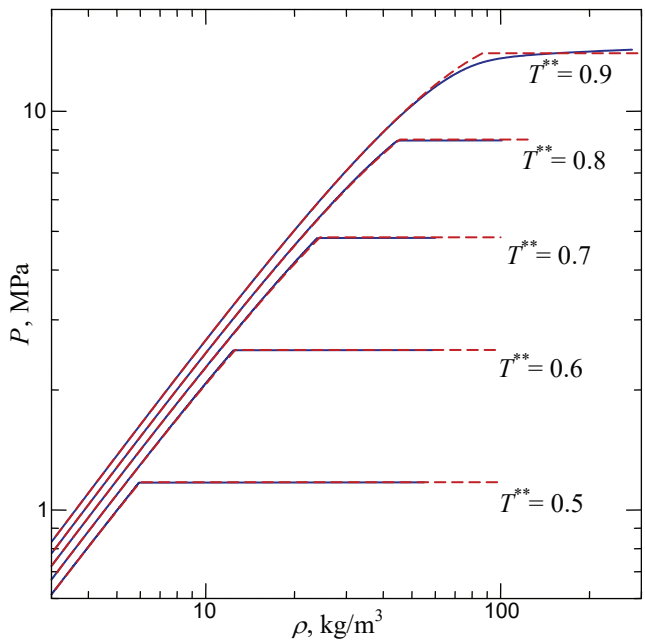}
\caption{\label{fig:3}Isotherms of water reproduced by Eq.~(\ref{eq:2}) with the set of 10000 reducible integrals, $\{ b_{n}^{emp} \}$, approximated in Eq.~(\ref{eq:6}) ($n_{0} = 500$) by using the information on the second virial coefficient and the phase-transition fugacity coefficient \cite{Wagner} at various reduced temperature ($ T^{**} = \frac { T - T_{tr} } { T_{cr} - T_{tr}} $, where $T_{tr} = 273.16$~K and $T_{cr} = 647.096$~K are the triple-point and critical temperatures, respectively). The dotted lines show the corresponding experimental isotherms \cite{Wagner}.}
\end{figure}

This feature allows application of the method to the real substances, for which only the experimental data on the second virial coefficient (and, sometimes, the third one) are known while the actual interaction potential remains unknown exactly. For example, the presented approach was applied to the water as a real substance with complex properties and non-spherical interaction potential surely distinct from the Lennard-Jones one. The results are partially shown in Fig.~\ref{fig:3} (actually, the saturation parameters vary by three-four orders (!) between the triple and critical points). Each isotherm of this Figure was reproduced by Eq.~(\ref{eq:2}) with the set of $10000$ reducible integrals approximated in Eq.~(\ref{eq:6}) exclusively on the basis of the second virial coefficient, $B_{2}$, and phase-transition fugacity, $ f $ (as the fugacity is the pressure of the ideal gas with the same chemical potential at the same temperature, the activity of Eq.~(\ref{eq:2}) is the particle number density of such ideal gas, $ z_{S}^{emp} = \frac {f} {{k_{B}} T} $). Both base quantities for each temperature were taken from the known experimental data \cite{Wagner}. 

Actually, all the considerations about the number of reducible integrals in Eq.~(\ref{eq:2}) or the $n_{0}$ parameter of Eq.~(\ref{eq:6}) stated above for the Lennard-Jones fluid remain correct for water as well. It is quite expected that the increase of temperature leads to the increase of deviations between the theoretical and real isotherms (at low temperatures, the results are even better than those presented in Fig.~\ref{fig:3}). Adding the third virial coefficient (also well known for water) neither improves nor worsens the results: at low temperatures, the main contribution to the isotherm curvature obviously belongs to the second virial coefficient (the saturation~-- the divergence of activity series due to the integrals of very high orders~-- begins long before the third, fourth or other relatively low-order integrals can significantly influence the curvature); at temperatures close to the critical one, the contribution of the much greater number of virial coefficients (irreducible integrals) becomes valuable and the third virial coefficient alone does not improve the accuracy.

Of course, the proposed method to reproduce the subcritical isotherms up to the saturation point is far from perfect. First of all, it is unable to reproduce the isotherms in condensed regimes as well as at the vicinity of the boiling point because the issue on the volume-dependence of cluster integrals is insufficiently explored at the moment \cite{PRE5} and remains a serious challenge for the modern cluster-based theory of condensation. Then we should remember that though the method is theoretical in essence it nevertheless uses some empirical data that may reduce its practical importance: it surely needs the further theoretical grounding and enhancement. On the other hand, this is the first time when a theoretical equation of state (actually, Eq.~(\ref{eq:2}) is a 100-years old equation rigorously grounded on the basis of statistical mechanics) reproduces the isotherms so close to the real ones qualitatively (without using any additional modification like the Maxwell construction) as well as quantitatively (especially for real substances with unknown interaction potentials). It should also be noted that, in all considered cases (the Lennard-Jones model as well as water), the initial information does not explicitly involve the empirical values of saturation density or pressure: only correct evaluation of low-order and high-order cluster integrals has first yielded so good agreement with experimental data. Such consistency between theory and practice must finally remove any doubts in the adequacy of Mayer`s cluster-based approach and, at the same time, considerably advances this theoretical approach in the direction of its practical usage.



\providecommand{\noopsort}[1]{}\providecommand{\singleletter}[1]{#1}%

\end{document}